\documentclass[letter,twocolumn]{jpsj3}
\usepackage{txfonts}

\title{Modulation Induced Phase Transition from Fractional Quantum Hall to Stripe State at $\mathbf{\nu=5/3}$}

\author{Akira \textsc{Endo}\thanks{E-mail address: akrendo@issp.u-tokyo.ac.jp}, Naokazu \textsc{Shibata}$^{1}$, and Yasuhiro \textsc{Iye}}

\inst{Institute for Solid State Physics, University of Tokyo, Kashiwa, Chiba 277-8581, Japan \\
$^{1}$Department of Physics, Tohoku University, Sendai, Miyagi 980-8578, Japan
}

\abst{We have investigated the effect of unidirectional periodic potential modulation on the fractional quantum Hall (FQH) state at filling factors $\nu=5/3$ and $4/3$. For large enough modulation amplitude, we find that the resistivity minimum at $\nu=5/3$ gives way to a peak that grows with decreasing temperature. Density matrix renormalization group calculation reveals that phase transition from FQH state to unidirectional striped state having a period $\sim 4 l$ (with $l$ the magnetic length) takes place at $\nu=1/3$ (equivalent to $\nu=5/3$ by the particle-hole symmetry) with the increase of the modulation amplitude, suggesting that the observed peak is the manifestation of the stripe phase.}

\kword{fractional quantum Hall effect, stripe phase, potential modulation, density matrix renormalization group, two-dimensional electron gas, unidirectional lateral superlattice}

\begin{document}
\maketitle

Strong interaction between electrons in a partially filled Landau level (LL) gives rise to varieties of exotic electronic states, including the fractional quantum Hall (FQH) state \cite{Tsui82,Perspectives97}, the stripe and bubble phases [unidirectional and two-dimensional hexagonal-lattice charge-density-wave (CDW) states]\cite{Fogler96,Moessner96}, and the Wigner crystal \cite{Perspectives97,Ye02,Chen03}. The state that wins out as the ground state switches from one state to another following the subtle change in the Coulomb interaction caused by the change in the applied magnetic field $B$, or the change in both the partial filling $\nu^*=\nu - [\nu]$ and the index $N=[\nu/2]$ of the topmost (partially filled) LL \cite{Shibata03}, where $\nu=n_\mathrm{e} h/eB$ is the LL filling factor with $n_\mathrm{e}$ the electron density and $[x]$ denotes the integer part of $x$. Note that $\nu^*$ is directly related to the inter-electron distance, while $N$, combined with the magnetic length $l=\sqrt{\hbar /(eB)}$, affects the interaction through the size and shape of the wave function. The FQH states have been found as the ground state only in the lowest ($N=0$) and the first excited ($N=1$) LLs, the LLs characterized by strong inter-electron Coulomb repulsion. The stripe or bubble phase becomes prevalent at higher LLs. There the wave function is spatially more extended and possesses nodes, resulting in the softening of the short-range repulsion, which favors the formation of electron clusters and hence the CDW states.
Experimentally, the FQH state is identified by vanishing longitudinal resistivity $\rho_{xx}=0$ and the plateau in the transverse resistivity $\rho_{yx}=h/e^2 \nu$ \cite{Perspectives97}. Strongly anisotropic resistivity \cite{Lilly99,Du99} and the reentrant integer quantum Hall effect \cite{Cooper99} discovered experimentally have been interpreted as the manifestation of the stripe and the bubble phases, respectively; the interpretation is corroborated by more recent observation of the microwave resonance \cite{Lewis02,Sambandamurthy08} ascribed to the pinning mode of the stripe or bubble phase.

The difference in energy between different phases is often very small, and transition between the phases can be made at fixed $\nu^*$ and $N$ by manipulating external parameters. For instance, $\nu=5/2$ and $7/2$ FQH states ($\nu^*=1/2$ at $N=1$ LL) were found to be replaced by the anisotropic state with the application of an in-plane magnetic field \cite{Pan99,Lilly99t,Rezayi00}. Similarly, the present authors showed that a weak unidirectional periodic potential modulation (having a period  $a=92$ nm close to that theoretically predicted for the stripe state), which is expected to be advantageous to the stripe phase, actually induces anisotropic resistivity at $\nu=5/2$ and $7/2$ \cite{Endo02f,Endo02I,Endo03lt}. So far, experimental evidence that can be related to the stripe or bubble phase has been limited to $N\geq1$ LLs, and has never been observed in the $N=0$ LL\@. Density matrix renormalization group (DMRG) calculation, however, predicts that the stripe phase, albeit with characteristics somewhat different from those in higher LLs \cite{Shibata03,Shibata04L,Shibata04}, does become the ground state also for the $N=0$ LL at $\nu^*\sim$ 0.42, 0.37 and between 0.32 and 0.15 (and also at the fillings equivalent to them by the particle-hole symmetry) \cite{Shibata03}.

In the present paper, we investigate the $N=0$ LL in unidirectional lateral superlattices (ULSLs) --- two-dimensional electron gases (2DEGs) subjected to unidirectional periodic potential modulation. We observe a peak, instead of the usual minimum, that grows with decreasing temperature at $\nu=5/3$ when the amplitude $V_0$ of the applied modulation is made sufficiently large. (The measurements are restricted to the range $1<\nu<2$ owing to the restriction in the available magnetic fields). We have performed DMRG calculation, which reveals phase transition from the FQH state to the stripe phase with the increase of $V_0$ and suggests that the observed peak reflects the modulation-induced stripe phase in the $N=0$ LL\@.

\begin{figure}[tb]
\includegraphics[bbllx=20,bblly=60,bburx=560,bbury=800,width=8.5cm]{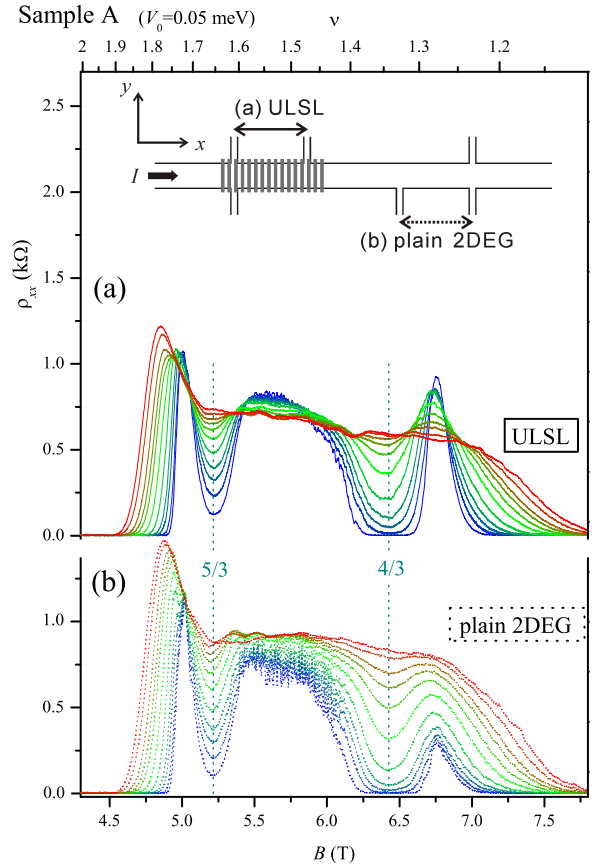}%
\caption{\label{d5001GF} (Color online) Magneto-resistivity traces ($1<\nu<2$) for the ULSL section (a) and the adjacent plain 2DEG section (b) of sample A, taken at temperatures $T$ (mK) = 15 (bottom, blue), 45, 70, 88, 120, 178, 226, 277, 334, 435, 604 (top, red). Inset: schematic diagram of the sample.}
\end{figure}

We examine two ULSL samples with different $V_0$ (sample A: $V_0=0.05$ meV, $a=115$ nm; sample B: $V_0=0.31$ meV, $a=184$ nm) fabricated from the same GaAs/AlGaAs 2DEG wafer with $n_\mathrm{e}=2.1 \times 10^{15}$ m$^{-2}$ and the mobility $\mu=70$ m$^2$/Vs. We introduced the modulation by employing the piezoelectric effect due to strain exerted on the wafer by a grating of resist placed on the surface \cite{Skuras97,Endo00e}. The simplicity of the fabrication procedure, along with the high resolution of the negative electron-beam resist used (calixarene) \cite{Fujita96}, allows us to introduce short period modulation with minimal disturbance or disorder in the coherence of the periodicity, which is indispensable to the present study. Unlike a more usual approach using a metallic grating, however, our method does not allow in-situ control of the modulation amplitude $V_0$. Noting that $V_0$ strongly depends on the period $a$ \cite{Endo05HH}, we prepared the two samples with differing $V_0$ by altering the periods. Each sample contains a section without modulation (plain 2DEG) in series for reference, as depicted in the inset of Fig.\ \ref{d5001GF}. Measurements were done in a dilution fridge by standard low-frequency (13 Hz) ac lock-in technique with a small current (0.5 nA) to avoid electron heating. 

\begin{figure}[tb]
\includegraphics[bbllx=20,bblly=60,bburx=560,bbury=800,width=8.5cm]{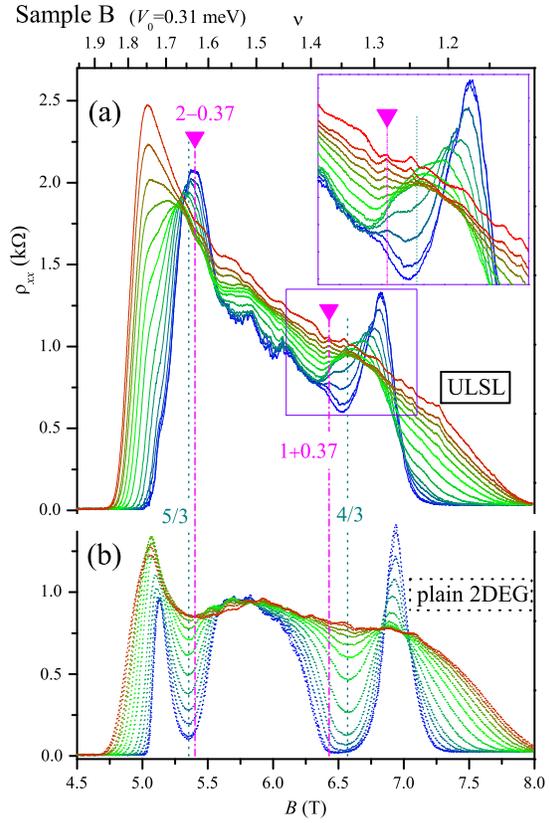}%
\caption{\label{5015d1GF} (Color online) Magneto-resistivity traces ($1<\nu<2$) for the ULSL section (a) and the adjacent plain 2DEG section (b) of sample B, taken at temperatures $T$ (mK) = 15 (bottom, blue), 25, 49, 70, 90, 117, 170, 218, 275, 323, 375, 434, 480 (top, red). Downward solid triangles with vertical dash-dotted lines mark the fillings at which the stripe correlation is predicted to be enhanced in the lowest Landau level \cite{Shibata03}. Inset: magnified view of the vicinity of $\nu=4/3$ in (a).}
\end{figure}

The main panels in Figs.\ \ref{d5001GF} and \ref{5015d1GF} show resistivity $\rho_{xx}$ of samples A and B, respectively, in the range $1<\nu<2$ for different temperatures $T$. As can be seen in Fig.\ \ref{d5001GF}, the modulation leaves $\rho_{xx}$ qualitatively unchanged in sample A\@. In marked contrast, the large amplitude modulation in sample B (Fig.\ \ref{5015d1GF}) drastically alters the $\rho_{xx}$ traces: the FQH minimum at $\nu=5/3$ that deepens with decreasing $T$ is replaced by a peak that grows with the cooling. By comparison, the $\nu=4/3$ FQH state appears to be less affected, retaining the residual minimum. The better stability against the modulation of the even-numerator FQH state is in line with our previous report \cite{Endo10FQHEMOD}. Closer look at $\nu\sim5/3$ reveals that the position of the peak in the ULSL shifts with decreasing $T$ toward $\nu=2-0.37$ (shown by dot-dashed line), the filling at which the enhancement of the stripe correlation is predicted \cite{Shibata03}. Also in the vicinity of $\nu\sim4/3$, a small hump develops at $\nu=1+0.37$ (dot-dashed line) at the lowest temperatures, as highlighted in the inset. These observations are consistent with the interpretation that the peak and hump in sample B are caused by the development of the modulation-induced stripe phase at low temperatures.

\begin{figure}[tb]
\includegraphics[bbllx=60,bblly=45,bburx=740,bbury=520,width=8.5cm]{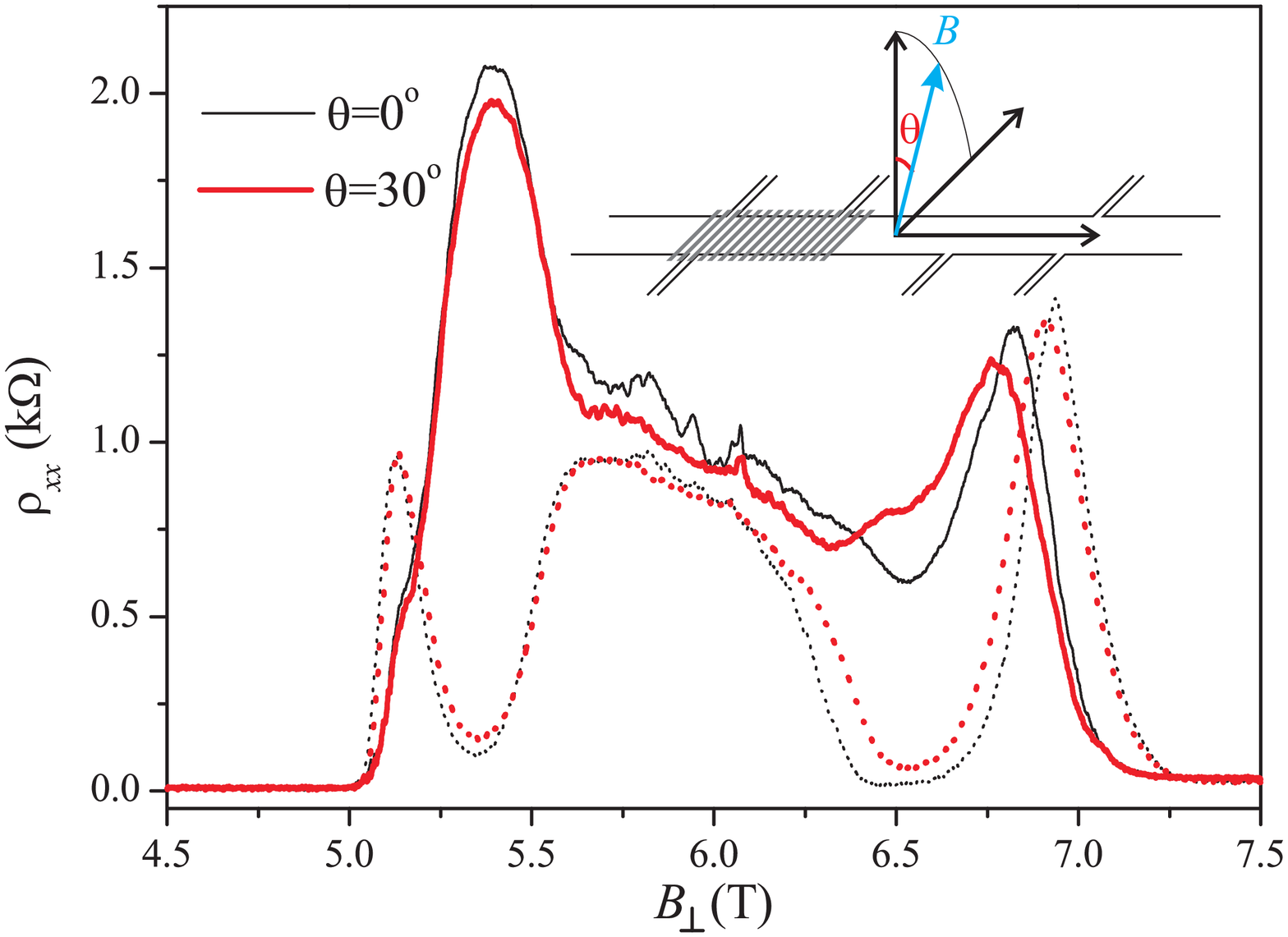}%
\caption{\label{R30rhoxxT} (Color online) Resistivity traces for tilt angle $\theta=0^\circ$ (thin, black) and 30$^\circ$ (thick, red) for the ULSL (solid curves) and the plain 2DEG (dotted curves) sections of sample B, plotted against $B_\perp=B \cos \theta$.}
\end{figure}

Further insight into the nature of the peak and hump can be inferred from our preliminary tilt-field experiment shown in Fig.\ \ref{R30rhoxxT}. The in-plane magnetic field $B_\|$ introduced by tilting is expected to destabilize the stripe phase when applied along the stripe as depicted in the inset of Fig.\ \ref{R30rhoxxT} \cite{Jungwirth99,Stanescu00}. The peak at $\nu\sim5/3$ is observed to slightly diminish by the tilting, consistent with the destabilization. More complicated behavior around $\nu\sim4/3$ can be interpreted in terms of the competition between the FQH and the stripe states.
The spin-unpolarized $\nu=4/3$ FQH state \cite{Clark89} weakens with the tilt, as is evident from the shrinking of the minimum in the plain 2DEG. The hump in the ULSL becomes more apparent with the tilting as the result of the weakening of the FQH state.
To fully clarify the effect of $B_\|$, however, we need to perform systematic studies varying the strength and the direction of $B_\|$.

\begin{figure}[tb]
\includegraphics[bbllx=65,bblly=30,bburx=730,bbury=530,width=8.5cm]{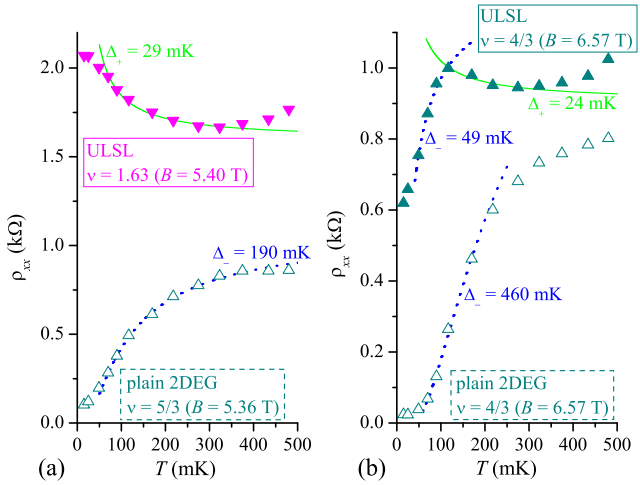}%
\caption{\label{TD5343} (Color online) Temperature dependence of $\rho_{xx}$ at fixed $B$ corresponding to the fillings $\nu$ noted in the figure for both the ULSL (solid symbols) and the plain 2DEG (open symbols) sections of sample B\@. The solid (dotted) curves are fits to Eq.\ (\ref{actv}) with the positive (negative) sign, with the value of $\Delta_\pm$ noted in the figure.}
\end{figure}

The intricate character of the state at $\nu\sim4/3$ in the ULSL is also reflected in the temperature dependence of the $\rho_{xx}$ plotted in Fig.\ \ref{TD5343}(b). Below $\sim 300$ mK, $\rho_{xx}$ increases with decreasing temperature, and then switches to downturn at still lower temperatures $T \le \sim 100$ mK\@. Both increasing and decreasing regimes can roughly be described by the activation-type temperature dependence,
\begin{equation}
\rho_{xx}=\rho_\infty \exp(\pm \Delta_\pm / 2 k_\mathrm{B} T),
\label{actv}
\end{equation}
with the positive and negative sign, respectively, as shown in Fig.\ \ref{TD5343}(b), except for the lowest temperature regime where the description by the variable-range hopping will be more appropriate \cite{Furlan98}. The behavior can be interpreted as the dominance of the stripe phase at relatively high temperatures, which is supplanted by the FQH state at the lower temperatures at $\nu = 4/3$. At slightly lower $B$ ($\nu=1.37$), the stripe phase continues to be the ground state down to the lowest temperature, as is revealed by the presence of the hump (Fig.\ \ref{5015d1GF}). The onset temperature $\sim 300$ mK of the rising of $\rho_{xx}$ both at $\nu=1.63$ [Fig.\ \ref{TD5343}(a)] and $4/3$ may be related to the onset of the development of the stripe phase. The values of $\Delta_+$ obtained by the fitting are difficult interpret at the present stage for lack of detailed knowledge as to how the stripe phase affects the resistivity. 

\begin{figure}[tb]
\includegraphics[bbllx=35,bblly=25,bburx=560,bbury=660,width=8.5cm]{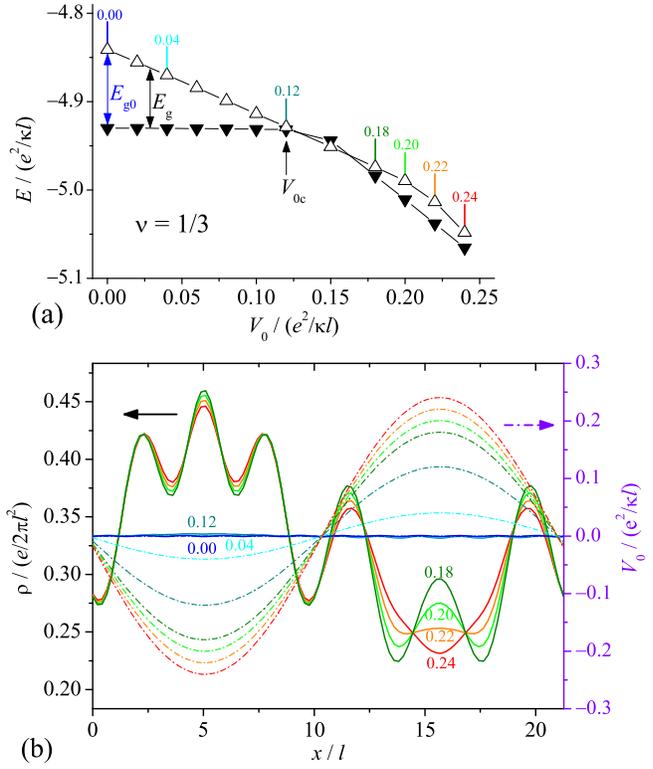}%
\caption{\label{EVnx} (Color online) (a) Dependence of the lowest two energy levels on the modulation amplitude $V_0$. (b) Electron charge density profiles $\rho (x)$ (in the unit of $e/2 \pi l^2$) for different values of $V_0$ noted in the figure (in the unit of $e^2/\kappa l$). Corresponding modulation profiles $V(x)$ are plotted by dot-dashed lines (with the corresponding colors, right axis). The values of $V_0$ are marked by the vertical ticks in (a). Both (a) and (b) are calculated by the DMRG method at $\nu=1/3$ for 12 electrons. See, e.g., Ref.\ \citen{Shibata01} for details of the DMRG method.}  
\end{figure}

We have performed DMRG calculations to explore the electronic states of the $N=0$ LL subjected to the modulated potential $V(x)=V_0 \cos(2 \pi x/a)$. The results for  $a=21.2 l$ (corresponding to 235 nm at $B=5.36$ T) at $\nu=1/3$, equivalent to $\nu=5/3=2-1/3$ by the particle-hole symmetry that takes account of the spins, are presented in Fig.\ \ref{EVnx}. As shown in Fig.\ \ref{EVnx} (a), the FQH gap $E_\mathrm{g}$ diminishes with increasing $V_0$ from $E_{\mathrm{g}0} \sim 0.09$ (in the unit of the Coulomb energy $e^2/\kappa l$) at $V_0=0$, until it vanishes at $V_{0\mathrm{c}} \sim 0.12$, namely when the modulation is roughly as large as the original FQH gap $E_{\mathrm{g}0}$. Figure \ref{EVnx} (b) reveals that the uniform electron density profile is abruptly, synchronized with the disappearance of the gap, transformed into the density wave having the wave length $\sim 4 l$ superposed on the slowly varying density undulation that follows the potential modulation; phase transition from the incompressible FQH state to the gapless stripe phase takes place with the increase of $V_0$. The charge density wave with the wave length $\sim 4 l$ is reminiscent of the stripe phase in higher LLs \cite{Fogler96,Moessner96,Shibata01}. We have also carried out calculations for different values of $a$ ranging from $10.6 l$ to $21.2 l$ \cite{Endoinprep}, and obtained qualitatively the same results. Notably, the value of the modulation amplitude $V_{0\mathrm{c}} \sim 0.12$ at which the gap vanishes does not vary with $a$.

\begin{figure}[tb]
\includegraphics[bbllx=60,bblly=25,bburx=800,bbury=540,width=8.5cm]{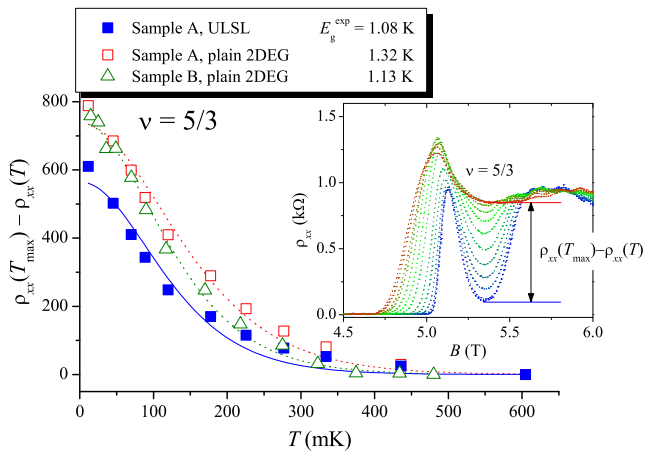}%
\caption{\label{d51Dgl} (Color online) Temperature dependence of the depth of the $\nu$=5/3 FQH minimum, defined in the inset, for sample A (squares) and B (triangles). ULSL and the plain 2DEG are plotted with solid and open symbols, respectively. Note that the minimum does not occur for the ULSL section in sample B\@. The curves are fits to Eq.\ (\ref{Dgl}) with the values of $E_g^\mathrm{exp}$ noted in the figure.}
\end{figure}

We now make an attempt to interpret our experimental results for the two samples in terms of the diagram Fig.\ \ref{EVnx} (a). First, we deduce the gap $E_\mathrm{g}^\mathrm{exp}$ of the $\nu=5/3$ FQH state from our experimental data by applying the Lifshitz-Kosevich formula \cite{Lifshitz56}, 
\begin{equation}
\Delta\rho_{xx}(T)=\rho_{xx}(T_\mathrm{max})-\rho_{xx}(T) \propto X_T / \sinh X_T
\label{Dgl} 
\end{equation}
with $X_T = 2 \pi^2 k_\mathrm{B} T / E_\mathrm{g}^\mathrm{exp}$, to the depth $\Delta\rho_{xx}(T)$ of the FQH minimum, which is regarded as the amplitude of the Shubnikov-de Haas oscillations of the composite fermions (CFs) \cite{Du94}. By the fitting shown in Fig.\ \ref{d51Dgl}, we obtain the gaps for the plain 2DEG sections in the both samples and also for the ULSL section in sample A, as noted on the top of Fig.\ \ref{d51Dgl}. Although $E_\mathrm{g}^\mathrm{exp}$ thus obtained, representing the energy difference between the centers of LLs (of CFs), is expected to be less vulnerable to the disorder compared with the activation gap [$\Delta_-$ in Eq.\ (\ref{actv})] \cite{Leadley97}, the values of $E_\mathrm{g}^\mathrm{exp}$ are still roughly an order of magnitude smaller than $E_\mathrm{g} \sim 0.1$ $e^2/\kappa l \sim 10$ K inferred from Fig.\ \ref{EVnx} (a). %
We note that the phase transition is provoked by the competition between the modulation $V_0$ and the original FQH gap $E_{\mathrm{g}0}$. In order to assess correctly from the diagram the effect of $V_0$ in our experimental devices, therefore, the energy scale in Fig.\ \ref{EVnx} (a) should be altered (in both horizontal and vertical axes, to preserve the decrement rate $\mathrm{d} E_\mathrm{g}/\mathrm{d} V_0$) so that $E_{\mathrm{g}0}$ equates with $E_\mathrm{g}^\mathrm{exp}$ of the plain 2DEG section. 
In the rescaled unit, the amplitudes of the modulation for samples A and B are translated to $V_0=0.040$ $(<V_{0\mathrm{c}})$ and $0.28$ $(>V_{0\mathrm{c}})$, respectively, in Fig.\ \ref{EVnx} (a). Thus, the diagram predicts slight reduction of the gap $E_\mathrm{g}$ by the modulation, $E_\mathrm{g} / E_{\mathrm{g}0} = 0.68$, for the ULSL section in sample A, which is in reasonable agreement with the experimentally obtained ratio $1.08/1.32=0.82$. Sample B, on the other hand, is located deep in the regime of the stripe phase, consistent with  the missing of the FQH minimum and the presence of a peak instead that can be related to the stripe phase.

The DMRG calculation is also performed on the spin-unpolarized $\nu=2/3$ FQH state (equivalent to $\nu=4/3$ by the particle-hole symmetry) \cite{Endoinprep}. The result predicts the collapse of the FQH gap at the value of $V_{0\mathrm{c}}$ similar to that in $\nu=1/3$, and therefore fails to account for the experimentally observed better robustness of the $\nu=4/3$ FQH state; the origin of the disparity is the subject of our future study. Interestingly, density wave with the wave length $\sim 4 l$ is found to be less prominent compared with the case in $\nu = 1/3$. This may explain, combined with the robustness the FQH state, the observed much smaller peak, actually a hump, at $\nu \sim 4/3$, which is in a remarkable contrast to the pronounced peak at $\nu \sim 5/3$. 

To summarize, we have examined the effect of unidirectional periodic potential modulation on the FQH states. We observe slight reduction of the $\nu = 5/3$ FQH gap for a small modulation amplitude $V_0$, and the collapse of the FQH state and the emergence of a prominent peak for a large $V_0$. The behavior is consistent with the modulation induced phase transition from the FQH to the stripe state predicted by the DMRG calculation, and provides experimental evidence for the presence of the stripe phase in the lowest LL\@.

\section*{Acknowledgment}
This work was supported by Grant-in-Aid for Scientific Research (C) (18540312) and (A) (18204029) from the Ministry of Education, Culture, Sports, Science and Technology.

\bibliography{qhe,twodeg,lsls,ninehlvs,ourpps,wc,notemdtrs}

\end{document}